\documentclass[]{spie}  

\usepackage{amsmath,amsfonts,amssymb}
\usepackage{graphicx}
\usepackage[colorlinks=true, allcolors=blue]{hyperref}
\usepackage[dvipsnames]{xcolor}
\usepackage{colortbl}
\usepackage{caption}
\usepackage{subcaption}
\newcommand{\gp}{GRAVITY$^+$}
\newcommand{\gw}{GRAVITY Wide}

\title{GRAVITY+ Wide: Towards hundreds of z $\sim$ 2 AGN}

\author[1]{Antonia Drescher}
\author[1]{Maximilian Fabricius}
\author[1]{Taro Shimizu}
\author[8]{Julien Woillez}
\author[9]{Pierre Bourget}
\author[1]{Felix Widmann}
\author[1]{Jingyi Shangguan}
\author[4]{Christian Straubmeier}
\author[4]{Matthew Horrobin}
\author[9]{Nicolas Schuhler}
\author[1]{Frank Eisenhauer}
\author[8]{Frederic Gont\'e}
\author[1]{Stefan Gillessen}
\author[1]{Thomas Ott}
\author[2]{Guy Perrin}
\author[2]{Thibaut Paumard}
\author[3]{Wolfgang Brandner}
\author[3]{Laura Kreidberg}
\author[5]{Karine Perraut}
\author[5]{Jean-Baptiste Le Bouquin}
\author[6,7]{Paulo Garcia}
\author[10]{Sebastian H\"onig}
\author[11]{Denis Defr\`ere}
\author[1]{Guillaume Bourdarot}
\author[1]{Helmut Feuchtgruber}
\author[1]{Reinhard Genzel}
\author[1]{Michael Hartl}
\author[1]{Frank Haussmann}
\author[1]{Dieter Lutz}
\author[1]{Nikhil More}
\author[1]{Christian Rau}
\author[1,12]{Jonas Sauter}
\author[1]{Sinem Uysal}
\author[1,12]{Patrick Wessely}
\author[1]{Ekkehard Wieprecht}
\author[1,13]{Lukas Wimmer}
\author[1]{Senol Yazici}
\author{GRAVITY+ Collaboration}

\affil[1]{Max-Planck-Institut f\"ur Extraterrestrische Physik (MPE), Gie\ss enbachstra\ss e, Garching bei M\"unchen, Germany}
\affil[2]{LESIA, Observatoire de Paris, PSL Research University, CNRS, Sorbonne Universit\'es, UPMC Univ. Paris 06, Univ. Paris Diderot, Sorbonne Paris Cit\'e, 92195 Meudon Cedex, France}
\affil[3]{Max-Planck-Institut f\"ur Astronomie, K\"onigstuhl 17, 69117 Heidelberg, Germany}
\affil[4]{1. Physikalisches Institut, Universit\"at zu K\"oln, Z\"ulpicher Str. 77, 50937 K\"oln, Germany}
\affil[5]{Univ. Grenoble Alpes, CNRS, IPAG, 38000 Grenoble, France}
\affil[6]{CENTRA and Universidade de Lisboa - Faculdade de Ciencias, Campo Grande, 1749-016 Lisboa, Portugal}
\affil[7]{Faculdade de Engenharia, Universidade do Porto, rua Dr. Roberto
Frias, 4200-465 Porto, Portugal}
\affil[8]{European Southern Observatory, Karl-Schwarzschild-Str. 2, 85748 Garching, Germany}
\affil[9]{European Southern Observatory, Casilla 19001, Santiago 19, Chile}
\affil[10]{School of Physics \& Astronomy, University of Southampton,
Southampton, SO17 1BJ, UK}
\affil[11]{Institute of Astronomy, KU Leuven, Celestijnenlaan 200D, B-3001,
Leuven, Belgium}
\affil[12]{Department of Physics, Technical University Munich, James-Franck-Strasse 1, 85748 Garching, Germany}
\affil[13]{I. Physikalisches Intitut, Justus-Liebig-University Giessen, Heinrich-Buff-Ring 16, 35392 Giessen, Germany}

\authorinfo{Further author information:\\A.Drescher.: E-mail: drescher@mpe.mpg.de
}

\begin{document} 
\maketitle

\begin{abstract}
As part of the \gp \ project, the near-infrared beam combiner GRAVITY and the VLTI are currently undergoing a series of significant upgrades to further improve the performance and sky coverage. The instrumental changes will be transformational, and for instance uniquely position GRAVITY to observe the broad line region of hundreds of Active Galactic Nuclei (AGN) at a redshift of two and higher. The increased sky coverage is achieved by enlarging the maximum angular separation between the celestial science object (SC) and the off-axis fringe tracking (FT) star from currently 2 arcseconds (arcsec) up to unprecedented 30 arcsec, limited by the atmospheric conditions. This was successfully demonstrated at the VLTI for the first time.  
\end{abstract}


\section{INTRODUCTION}
\label{sec:intro}  

In its first five years of science operation, GRAVITY \cite{2017A&A...602A..94G} at the VLTI has delivered groundbreaking results in different areas in modern astronomy. In order to stay at the forefront of discovery and to go even further, a new combination of upgrades to the VLTI and GRAVITY is currently being implemented. This project is called \gp. \gp \ aims for all-sky and high contrast interferometry with milli-arcsec resolution imaging and faint science with a limiting magnitude of m$_K=22$\cite{Eisenhauer2019}. \gp \ is implemented in several phases, where some of them are already concluded, such as the implementation of two new grisms in 2019 to increase the instrument throughput by a factor $>2$ \cite{Yazici2021}. Further phases which are currently under development are the implementation of new deformable mirrors and state-of-the-art adaptive optics (AO) wavefront sensors, laser guide stars (LGS) on all four 8m UTs, an improved vibration control, and the reduction of existing noise sources in GRAVITY. For the latter, a new observing mode has been developed which suppresses the noise of the metrology laser and is called GRAVITY faint (F. Widmann, GRAVITY+ Collaboration et al. 2022 in prep.). Finally, \gp \ includes the implementation of wide-field off-axis fringe tracking, called \gw \cite{gwide_paper}, which is the topic of this paper. \gw \ allows for increasing the separation between the SC and FT from currently 2 arcsec up to about 30 arcsec. Together with the new AO system and LGS on all four UTs, \gp \ will have a significantly improved sky coverage, and will open up new science on faint objects, such as faint AGN out to redshift $\approx2$. \gw \ is implemented in two steps with the first step now concluded. The new observing mode is already commissioned and available for the community.

\begin{figure}[t]
\centering
\includegraphics[width=0.99\textwidth]{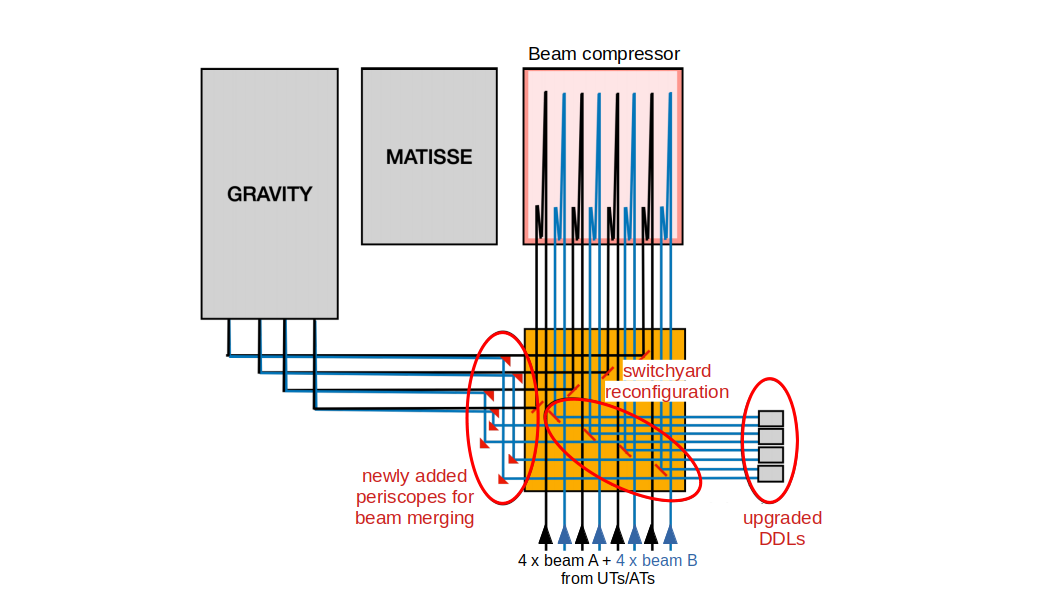}
\caption{Schematic overview of the VLTI switchyard and beam routing towards GRAVITY. In total, eight beams from four telescopes arrive at the switchyward: four beams dedicated to the FT target (beam A, black) and four beams dedicated to the SC target (beam B, blue). The figure shows the configuration for \gw \ observations with the periscopes in "IN" position. The switchyward reconfiguration relays the B beams coming from the beam compressors via the upgraded PRIMA DDLs to the periscopes to merge them into the A beams.}
\label{fig:schematic-switchyard} 
\end{figure}

\section{THE GRAVITY WIDE IMPLEMENTATION}
\label{sec:implementation}
In its dual-field observing mode, GRAVITY simultaneously observes a science object and a nearby fringe tracking star as a phase reference and to allow for long coherent integrations on the science object \cite{2017A&A...602A..94G}. The FT star has a limiting magnitude of m$_H=10$ and has to be in the same field-of-view as the SC object, which is 2 arcsec for the UTs, and 4 arcsec for the ATs. This requirement strongly limits the sky coverage of GRAVITY. With the new observing mode \gw \ we overcome this requirement by increasing the FT--SC separation from 2 arcsec up to several 10 arcsec. This allows to pick a FT star from a much larger area, thus observe faint targets further away from the galactic plane. The \gw \ implementation was done by reviving the existing dual-beam infrastructure at the VLTI, i.e. the star separators\cite{star_separator_prima} at the Coud\'e focus of each UT, the dual-beam capabilities of the main delay lines and the PRIMA differential delay lines\cite{espri_ddls} (DDLs). For a schematic overview see Fig.~\ref{fig:schematic-switchyard}. While the star separators and the dual-beam capabilities of the main delay lines could be used unchanged, the DDLs had to be slightly modified to properly relay the exit pupils towards GRAVITY. Additionally, the VLTI switchyard had to be upgraded. In \gw, the star separators are used to pick up the light from the FT and SC from a 2 arcmin field, which is then propagated separately to the VLTI lab. With now eight beams arriving at the switchyard, it had to be extended to route the four off-axis beams (B beams, blue in Fig.~\ref{fig:schematic-switchyard}) to the SC detector, and the four on-axis (A beams, black in Fig.~\ref{fig:schematic-switchyard}) to the FT detector.

\begin{figure}[h]
	\centering
	\includegraphics[width=0.5\textwidth]{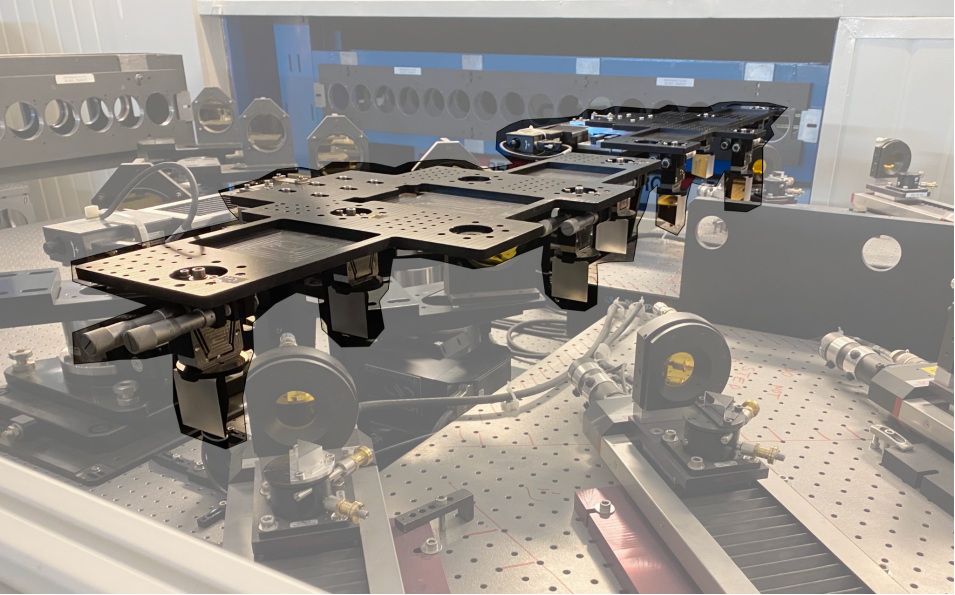}
	\caption{Newly added periscopes on the VLTI switchyward. They are implemented through flat mirrors that hang upside down from two motorized bridges to switch \gw \ on and off.}
	\label{fig:newly_added_periscopes} 
\end{figure}

Before the light is fed into GRAVITY, the A and B beams are merged with a 2 arcsec offset angle through the addition of four periscopes. The periscopes are implemented through eight flat mirrors that hang upside down from two bridge-like structures (see Fig.~\ref{fig:newly_added_periscopes}), which are motorized and can be moved in and out of the light path. Together with the DDLs they were shipped to Paranal in October 2021 and installed on the optical bench. In December 2021, sighting scopes were installed on the reference plate around the optical switchyard for alignment.

\begin{figure}[h]
	\centering
	\includegraphics[width=0.6\textwidth]{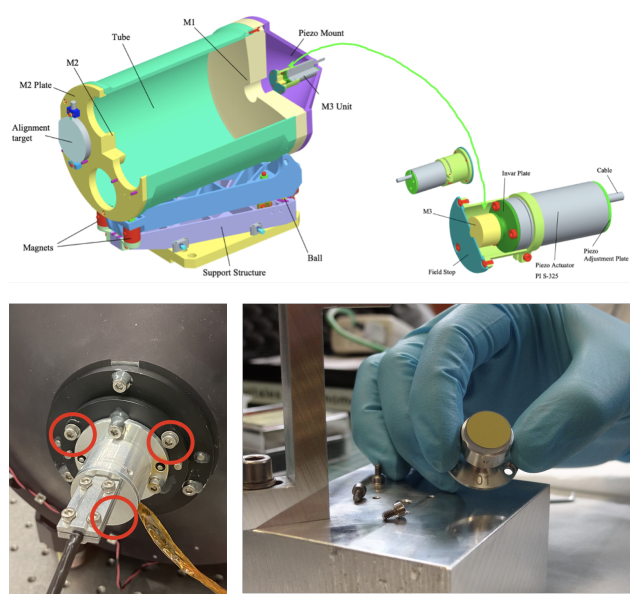}
	\caption{Replacement of the M3 mirrors of the PRIMA DDLs. The top figure shows a drawing of the PRIMA DDLs and the location of the M3 mirrors. The bottom left picture shows the piezo driven tip/tilt assembly that is accessible from the backside of the PRIMA DDLs. The red circles indicate the shim locations. The bottom right picture shows the newly fabricated M3 in the thermally clamped mount (here for a prototype, the final mounts were anodized black).}
	\label{fig:ddl_m3_replacement} 
\end{figure}

To compensate for the differential optical path difference (dOPD) between the FT (B beams) and SC (A beams), we use the PRIMA DDLs which are passed by the B beams. Four of the already existing PRIMA DDLS were modified to relay the pupil to the appropriate location taking the additional pathlength through the periscopes into account. We replaced the DDL-M3 mirrors with new mirrors with appropriately modified radii of curvature
%
(see Fig.~\ref{fig:ddl_m3_replacement} top and bottom right). 
The M3 are held in place also by thermal clamping, therefore, we needed new mirror mounts. The mounts at the backside of the DDL housings were removed and we replaced the mirror/mount package as a whole. Further, for realigning the lateral pupil location, reshimming of the M3 mount positions was required (see Fig.~\ref{fig:ddl_m3_replacement} bottom left). The alignment was done with the help of sighting scopes.
In the end, when the switchyard and periscopes were aligned we checked the pupil stability of the propagated beams through the DDLs. We found that no drifts in pupil position were observed.
\begin{figure}[h]
	\centering
	\includegraphics[width=0.5\textwidth]{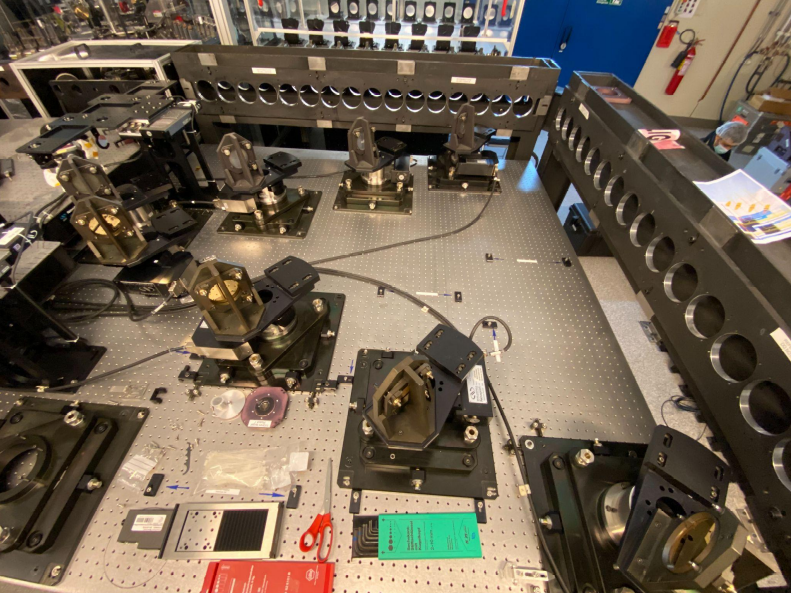}
	\caption{Picture of the reconfigured switchyard fold mirrors.}
	\label{fig:switchyard_reconfiguration} 
\end{figure}

Finally, the VLTI switchyard itself needed to be reconfigured to allow for the new beam routing (see Fig.~\ref{fig:switchyard_reconfiguration}). To relay the B beams from the beam compressors through the DDLs four of the switchyard mirrors were repositioned. In preparation for the second step of the \gw \ implementation and repeated repositioning of the mirrors, reference blocks were placed onto the optical bench.

\section{THE GRAVITY WIDE OBSERVING MODE}
\label{sec:observingmode}
In P110, \gw \ will be offered with the following options:
\begin{itemize}
    \item \textbf{Program type:} \gw \ will not be available for large programs and monitoring programs, i.e. $<100$ hours. However, this is expected to change in the next period.
    \item \textbf{Observing mode:} As for the observing modes, service mode is recommended, and visitor mode is possible.
    \item \textbf{Observation type:} The default type of observation will be snapshot. Additionally, time series, for observations with a critical time link in between, will be available. In contrast, imaging and astrometry will not yet be offered.
    \item \textbf{Arrays:} \gw \ can be used for both UTs and ATs, but only astrometric and small for the ATs.
    \item \textbf{Calibrations:} Broad band visibilities will not be available, which means SCI only.
    \item \textbf{Adaptive optics:} NAOMI will be available on the ATs, and MACAO on the UTs.
    \item \textbf{Spectral resolution:} HIGH and MEDIUM resolution observations will be offered.
    \item \textbf{Polarization:} Both SPLIT and COMBINED are possible. COMBINED is recommended for sensitivity.
    \item \textbf{Observables:} Differential visibilities and phases, and closure phases are offered observables. Absolute phases are not available, since they need the laser metrology system which is not yet implemented for \gw.
    \item \textbf{Limiting magnitudes and sensitivity:} Limiting magnitude and magnitude for preliminary sensitivity for one hour on target are given in the table below. The limitations are given for the 10$\%$ and 30$\%$ best atmospheric turbulence conditions, T. For worse atmospheric conditions, \gw \ is not offered.
    \begin{center}
    \begin{tabular}{ c|c|c|c|c}
    \hline \hline
     & &  T$<10\%$ & T$<30\%$ & \\
     \hline
    \gw \ on UTs & FT & 10.5m & 10.0m & \\
    & SC & 17.0m & 16.0m & H$_{SC}<$17 \\
    \hline
    \gw \ on ATs & FT & 9.5m & 9.0m & \\
    & SC & 14.0m & 13.0m & H$_{SC}<$14 \\
    \hline
    \end{tabular}
    \end{center}
    \item \textbf{Overheads:} The minimum sequence is 1800 s, thus 30 minutes, based on dual$\_$wide$\_$acq of 900 s and dual$\_$obs of 900 s\cite{overheads}.
\end{itemize}

\section{ATMOSPHERIC COHERENCE LOSS}
\label{sec:onskytests}
\begin{figure}[h]
	\centering
	\includegraphics[width=0.9\textwidth]{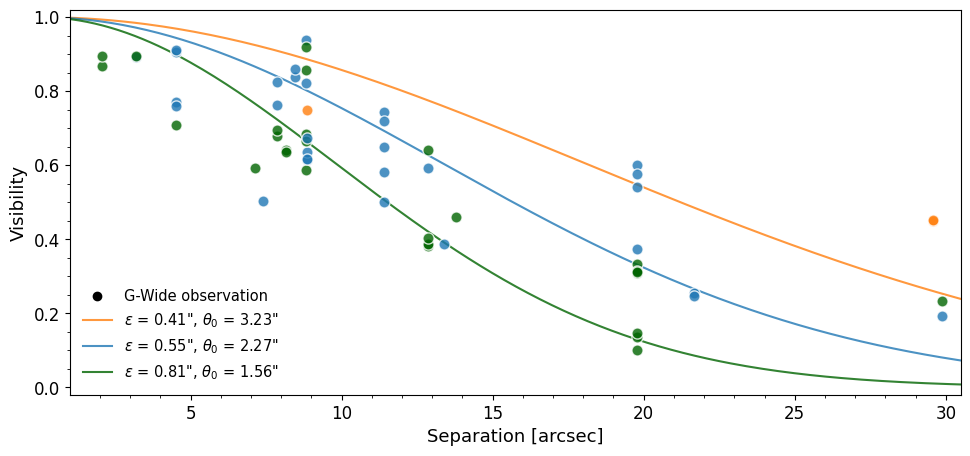}
	\caption{Observed contrast loss versus FT--SC separation. The three curves indicate the contrast loss for different seeing and isoplanatic angle values as expected from atmospheric turbulence following Ref. \citenum{elhalkouj2008} for the 1.8m ATs. Both the seeing, $\epsilon$, and the isoplanatic angle, $\theta_0$, are defined at 500 nm.}
	\label{fig:visloss_all_gwide} 
\end{figure}

During \gw \ commissioning, we observed several star pairs with the ATs with FT--SC separations ranging from 2 to 31 arcsec. Investigating and understanding the influence of atmospheric turbulence on the new large separation observing mode is of great importance, as the atmosphere leads to a coherence loss, especially at large separations. The goal is to find out up to which FT--SC separation it is possible to observe with \gw \ to not lose too much coherence. For this, we show the mean visibility of all six baselines for the \gw \ observations versus the FT--SC separation in Fig.~\ref{fig:visloss_all_gwide}. The
60 \gw \ observations are split into three groups, depending on the isoplanatic angle, $\theta_0$, measured during the exposure. The orange data points have the highest isoplanatic angle values with 3.12 arcsec $< \theta_0 <$ 3.3 arcsec, the blue data points span values of  2.02 arcsec $< \theta_0 <$ 2.7 arcsec, and the green data points correspond to the smallest isoplanatic angle values with 1.23 arcsec $< \theta_0 <$ 1.93 arcsec. We calculate the mean seeing and mean isoplanatic angle of each group, and compute the visibility loss as prediced by Ref.~\citenum{elhalkouj2008}:

\begin{equation}\label{equ:vis_loss}
V_{\text{average}}(\theta) = V \ \text{exp} \Big[ - \frac{2\pi^2}{\lambda^2}\sigma_p^2(\theta) \Big] \ .
\end{equation}
The isopistonic angle error, $\sigma_p (\theta)$, in Eq. (\ref{equ:vis_loss}) is given by:
\begin{equation}\label{equ:isopist_angle_error}
\sigma_p (\theta) \sim 0.12 \pi^{1/3} \lambda \left( \frac{D}{r_0} \right) ^{-1/6} \frac{\theta}{\theta_0} \hspace{1cm}  \text{for} \ L_0 \rightarrow \infty \ ,
\end{equation}

where $\lambda$ is the wavelength, $D$ the aperture diameter, $\theta$ the FT--SC separation, $r_0$ the Fried parameter with $r_0 = 0.98(\lambda / \epsilon)$ \cite{seeing}, where $\epsilon$ is the seeing, and $\theta_0$ the isoplanatic angle with $\theta_0 = 0.314(cos z)(r_0 / H)$ \cite{isoplanatic_angle}, where $z$ is the zenith angle, and $H$ the height of the turbulent layer.
The visibility loss for the three groups of $\theta_0$ given above is computed with Eq. (\ref{equ:vis_loss}) and represented by the three solid curves in Fig.~\ref{fig:visloss_all_gwide}. We find the contrast loss with increased separation as expected by the model by Ref.~\citenum{elhalkouj2008} from atmospheric anisoplanatism for an infinite outer scale of turbulence, $L_0$.
The contrast loss model depends on four parameters: the separation between SC and FT, the aperture diameter, the seeing, and the isoplanatic angle. We find that the expected visibility is higher for larger isoplanatic angles. This is because a larger isoplanatic angle means a lower turbulent layer, thus, a larger overlap of the projected pupils on sky. This leads to a better correction of wavefront aberrations by AO. We also find that the model is more
sensitive to the isoplanatic angle than to the seeing. We therefore proposed to extend the seeing categories for observing proposal preparation by this parameter.

We see that the coherence loss we observe with \gw \ overall matches the predictions from the model by Ref.~\citenum{elhalkouj2008}. A reason for a mismatch between some data points and the model could be that the model assumes an infinite outer scale of turbulence. However, this scale is about 22 m in the atmosphere model of Paranal \cite{outerscale}. This could lead to an underestimation of the maximum visibility in the model by Ref.~\citenum{elhalkouj2008}. Additionally, the contrast loss model does not take parameters such as DIT, total exposure time, coherence time, airmass or magnitude of the SC into account. Shorter DITs could result in higher visibilities, because they better catch the rapid change of atmospheric turbulence, or in other words, are short enough to "freeze" the turbulence.

\section{OUTLOOK}
\label{sec:outlook}
The second and last step of the \gw \ implementation is the integration of the opto-mechanical functionality of the PRIMA DDLs in the beam compressors. This means that the OPD control will then be handled by the upgraded beam compressors instead of the DDLs. The reconfiguration of the optical switchyard from step 1 of the \gw \ upgrade (described above) will mostly be undone. With the second step of the \gw \ implementation five optical reflections per B beam will be removed, which increases the optical throughput by expected $\approx15-20\%$. As a consequence, observations of even fainter science targets in a shorter amount of time will become possible. 

\section{CONCLUSION}
\label{sec:conclusion}
In this paper, we present GRAVITY's new large-separation fringe tracking mode, called \gw, which is part of the \gp \ project. \gw \ tremendously improves the sky coverage of GRAVITY by increasing the separation between the SC and FT from 2 arcsec up to 30 arcsec, limited by the atmospheric conditions. This makes it possible to observe faint objects further away from the galactic plane, thus open up the sky for even fainter and all-sky interferometry. We investigate the influence of atmospheric turbulence on the new observing mode and show that the coherence loss depends on the atmospheric parameters seeing and isoplanatic angle. We demonstrate that a larger isoplanatic angle leads to a higher visibility, therefore propose to take $\theta_0$ into account when preparing observing proposals. \gw \ is successfully commissioned, and available to the community.

 

\bibliography{report} 
\bibliographystyle{spiebib} 

\end{document}